\documentclass[twocolumn,secnumarabic,nobibnotes,reprint,superscriptaddress,amssymb,amsmath,aps,showpacs,10pt,floatfix,prx,longbibliography]{revtex4-2}

\usepackage{ulem}
\usepackage{graphicx}
\usepackage{amsfonts}
\usepackage{stmaryrd}
\usepackage{bbm}
\usepackage{mathrsfs}
\usepackage{tipa}
\usepackage{txfonts}
\usepackage{dcolumn}
\usepackage{epstopdf}
\usepackage{float}
\usepackage{chngcntr}
\usepackage{array}
\usepackage{longtable}
\usepackage{subfigure}
\usepackage[utf8]{inputenc}
\usepackage{newunicodechar}
\usepackage{titlesec} 

\usepackage[defaultcolor=red]{changes}
\usepackage{booktabs}
\usepackage{makecell}
\usepackage{multirow}

\usepackage[colorlinks,bookmarks=true,citecolor=blue,linkcolor=blue,urlcolor=blue]{hyperref} 

\makeatletter
\renewcommand{\thesubfigure}{\alph{subfigure}}
\renewcommand{\@thesubfigure}{(\thesubfigure)\hskip\subfiglabelskip}
\makeatother

\makeatletter
\DeclareRobustCommand{\delaycite}[1]{%
	\begingroup
	\let\if@filesw\iffalse 
	\cite{#1}%
	\endgroup}
\makeatother

\begin{document}

\title{Ultrafast Terahertz and Optical Spectroscopy under Synergetic Extreme Conditions}

\author{Xinbo Wang}
\email{xinbowang@iphy.ac.cn}
\affiliation{Beijing National Laboratory for Condensed Matter Physics, Institute of Physics, Chinese Academy of Sciences, Beijing 100190, China}
\affiliation{School of Physical Sciences, University of Chinese Academy of Sciences, Beijing 100190, China}

\author{Tao Dong}
\affiliation{Tsung-Dao Lee Institute, Department of Physics and Astronomy, Shanghai Jiao Tong University}
\affiliation{International Center for Quantum Materials, School of Physics, Peking University, Beijing 100871, China}

\author{Jianlin Luo}
\affiliation{Beijing National Laboratory for Condensed Matter Physics, Institute of Physics, Chinese Academy of Sciences, Beijing 100190, China}
\affiliation{School of Physical Sciences, University of Chinese Academy of Sciences, Beijing 100190, China}

\author{Nanlin Wang}
\affiliation{Tsung-Dao Lee Institute, Department of Physics and Astronomy, Shanghai Jiao Tong University}
\affiliation{International Center for Quantum Materials, School of Physics, Peking University, Beijing 100871, China}

\date{\today}

\begin{abstract}
Elucidating and manipulating emergent phases in complex materials requires direct access to their low-energy collective modes. Terahertz (THz) time-domain and ultrafast optical spectroscopies have emerged as indispensable experimental tools, enabling the probing of intrinsic electrodynamics and the coherent control of non-equilibrium states. At the Synergetic Extreme Condition User Facility (SECUF), we have developed a suite of intense ultrashort light sources covering the near-infrared, mid-infrared, and THz spectral ranges. By integrating these strong-field pulses with extreme sample environments, such as low temperatures, strong magnetic fields, and high pressures, we have established several state-of-the-art spectroscopy platforms. In this article, we outline the technical specifications of each setup and highlight representative user experiments. The presented results underscore the exceptional capability of the THz experimental unit (A4-2) to explore ultrafast dynamics across a multi-parameter thermodynamic phase space.
\end{abstract}

\maketitle

\section{Introduction}

Terahertz (THz) radiation bridges the spectral gap between microwaves and infrared light, typically ranging from 0.1 to 30 THz (1~THz $\approx$ 4.1~meV) within the electromagnetic spectrum. In condensed matter physics, the characteristic energy scales of elementary excitations, such as phonons, magnons, excitons, and superconducting gaps, fall within the THz and mid-infrared (MIR) requency regimes~\cite{Basov2011,UlbrichtRevModPhys2011,kampfrath2013resonant,Dong2023}. Terahertz time-domain spectroscopy (THz-TDS) provides a unique phase-resolved experimental tool to directly probe the linear electrodynamic response of these low-energy modes~\cite{Neu2018}. Over the past decade, the study of light-matter interactions has experienced a profound paradigm shift from weak-field probing to strong-field coherent control~\cite{Basov2017,deLaTorre2021}. Intense THz and MIR pulses can resonantly pump specific collective excitations, driving quantum materials far from equilibrium to trigger pronounced nonlinear optical effects and exotic transient quantum phase transitions, spanning from light-induced superconductivity~\cite{Fausti2011, Rowe2023, Fava2024} to ultrafast coherent magnetic switching~\cite{Schlauderer2019, Zhang2024}.

While intense light pulses drive quantum materials far from equilibrium, the resulting transient dynamics are dictated by their initial thermodynamic ground states~\cite{Basov2017,deLaTorre2021}. Applying extreme experimental conditions, such as low temperatures, strong magnetic fields, and high pressures, enables the systematic investigation of complex phase diagrams across diverse quantum materials~\cite{Tokura2017, Mao2018}. For instance, an external magnetic field provides a clean tuning knob to manipulate spin configurations and electronic topologies, whereas hydrostatic pressure continuously modulates the electronic bandwidth and electron-lattice interactions without introducing chemical disorder~\cite{Yamamoto2015}. Integrating ultrafast optical spectroscopies with such demanding environments constitutes a major frontier in experimental physics~\cite{Dong2023}. However, achieving this synergetic integration encounters formidable technical challenges, particularly for intense THz and MIR techniques~\cite{Leitenstorfer2023}. The difficulties stem from the inherent complexities in generating intense long-wavelength pulses and efficiently coupling them into the spatially restricted optical accesses of extreme sample environments~\cite{Wang2024RSI}.

\begin{figure*}
	\includegraphics[width=0.85\linewidth]{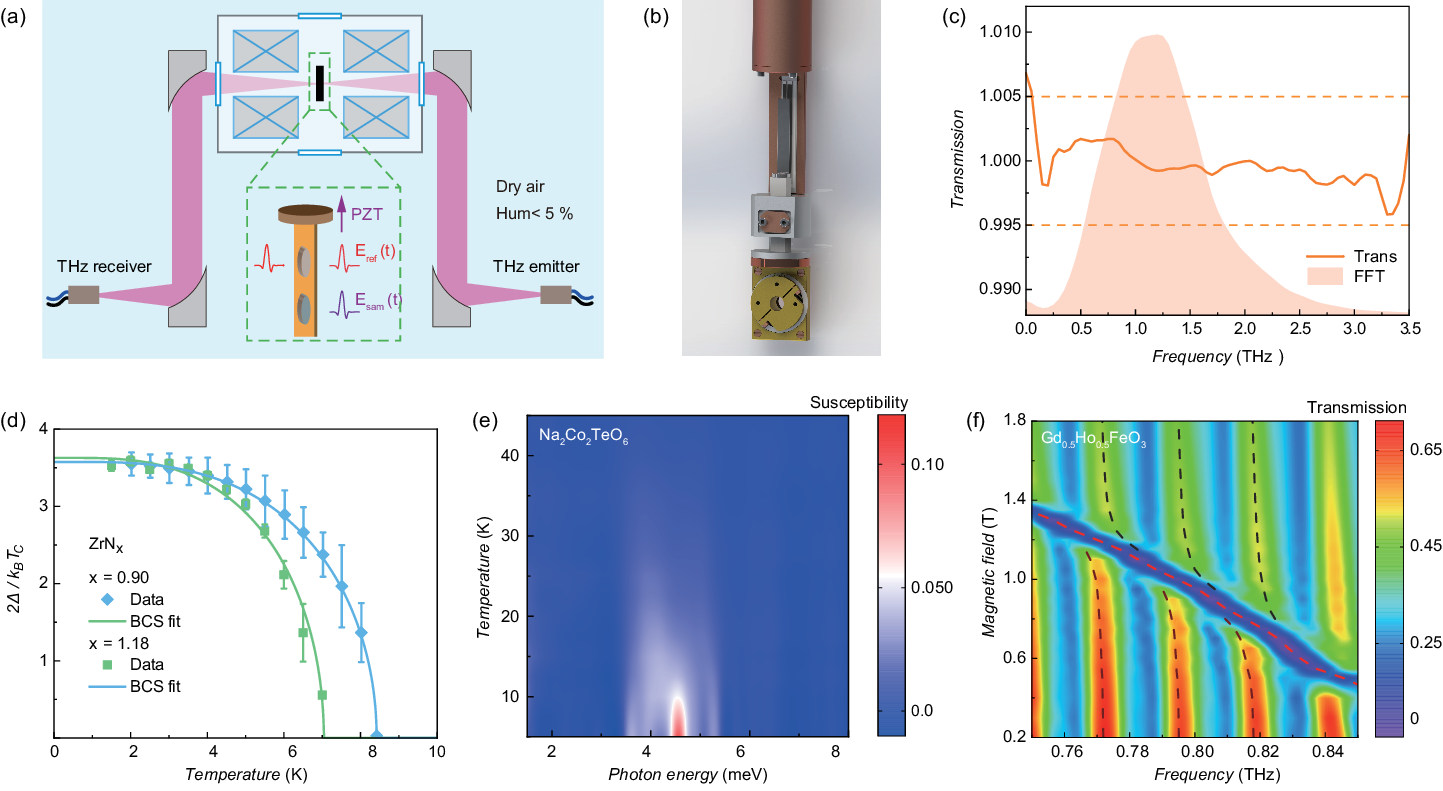}
	\centering		\vspace{-0.2cm}
	\subfigure{\label{1a}} 
	\subfigure{\label{1b}}
	\subfigure{\label{1c}}
	\subfigure{\label{1d}} 
	\subfigure{\label{1e}}
	\subfigure{\label{1f}} 
\caption{
	(a) Schematic illustration of the THz spectrometer coupled with a split-coil superconducting magnet. 
	(b) Detailed view of the PZT-driven linear and rotation stages mounted at the end of the sample rod. In THz-TDS measurements, the rotation stage is replaced by the customized sample holder shown in (a).  Adapted with permission from Ref.~\delaycite{Wang2024RSI}. Copyright 2024 AIP Publishing. 
	(c) Vacuum-to-vacuum transmittance of two 2.5-mm-diameter apertures. 
	(d) Temperature dependence of the derived superconducting energy gap for the disordered ZrN$_x$ films, with solid lines representing the standard BCS fit. Adapted with permission from Ref.~\delaycite{Chen2023ZrN}. Copyright 2023 Science China Press. 
	(e) Color map of the imaginary part of the magnetic susceptibility for the Kitaev quantum spin liquid candidate Na$_2$Co$_2$TeO$_6$, measured under a 10~T in-plane magnetic field. Adapted with permission from Ref.~\delaycite{Shi2025Na2Co2TeO6}, Copyright 2025 by the American Physical Society. 
	(f) Magnetic field-dependent THz transmittance contour plot of $b$-cut Gd$_{0.5}$Ho$_{0.5}$FeO$_3$ single crystal at 2~K. Adapted with permission from Ref.~\delaycite{Chen2025GdHoFeO3}. Copyright 2025 AIP Publishing.
}
\label{1}

\end{figure*}

In 2018, the initial design for the infrared and THz experimental stations at the Synergetic Extreme Condition User Facility was proposed~\cite{Dong2018CPB}. Following four years of successful commissioning, this article presents the established spectroscopic capabilities of the operational THz experimental unit (A4-2). We first introduce the THz time-domain spectroscopy coupled with low temperatures and strong magnetic fields. Next, we detail the generation of intense THz and mid-infrared sources and their associated pump-probe platforms, including THz-pump optical-probe techniques, THz third-harmonic generation, and THz two-dimensional coherent spectroscopy. We highlight the key scientific advances achieved by users on these respective platforms. Furthermore, we discuss the optical pump-probe spectroscopy integrated with high-pressure diamond anvil cells. Finally, we conclude this article with a perspective on future opportunities.

\section{Ultrafast Terahertz and Optical Spectroscopy Platforms}
\subsection{Terahertz Time-Domain Spectroscopy}
Probing energy gaps and collective modes in complex materials requires precise measurements of the low-energy electrodynamics~\cite{Basov2011, Neu2018}. To investigate the linear responses under synergetic extreme conditions, we have integrated a fiber-coupled THz spectrometer based on photoconductive antennas with a split-coil superconducting magnet system (Fig. \ref{1a}). The platform is equipped with a variable temperature insert to achieve low temperatures down to 1.5~K alongside magnetic fields up to 10~T. The entire magnet assembly is mounted on a linear stage and a swing bearing, permitting a 90$^{\circ}$ rotation of the magnetic field relative to the terahertz propagation direction~\cite{Wang2024RSI}. The mechanical design guarantees a convenient transition between the Faraday and Voigt configurations without altering the optical alignment. Furthermore, a piezoelectric (PZT)-driven linear stage with a travel range of 20~mm is installed at the end of the sample rod for precise vertical translation(Fig. \ref{1b}). It enables automated \textit{in situ} switching between the sample and reference apertures at low temperatures and high magnetic fields, thereby substantially enhancing measurement reliability and data acquisition efficiency.

\begin{figure*}
	\includegraphics[width=0.9\linewidth]{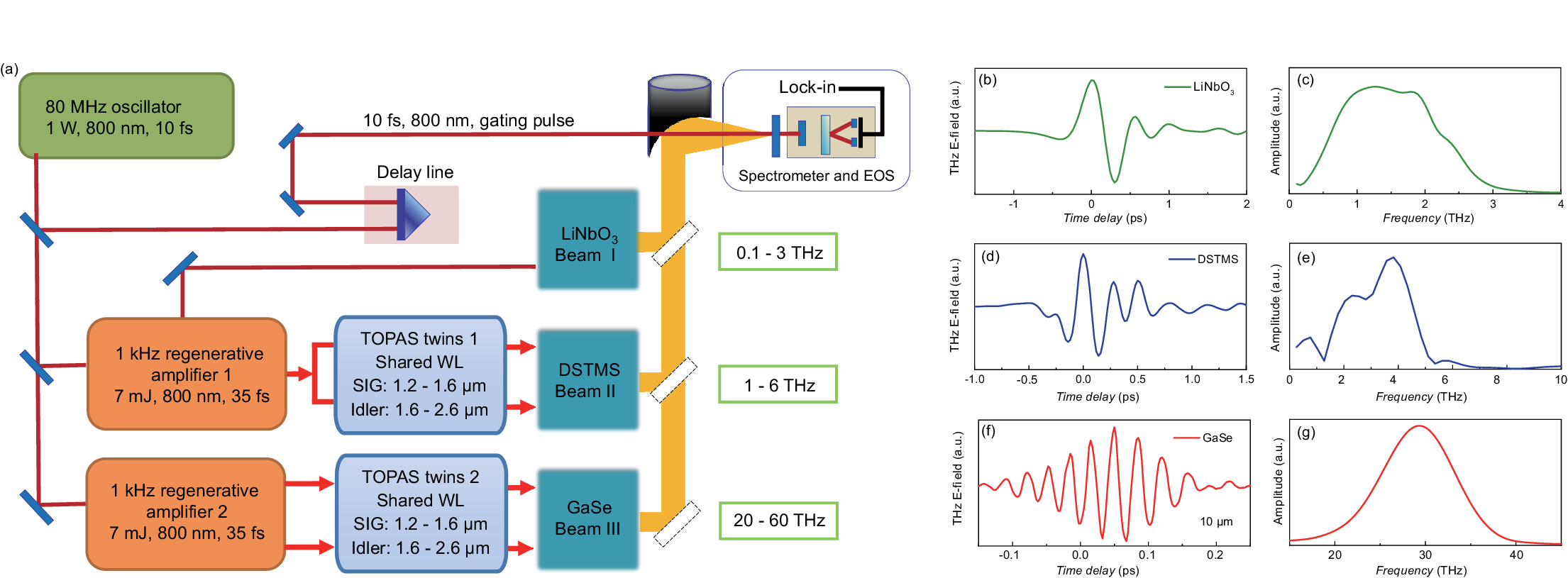}
	\centering		\vspace{-0.2cm}
	\subfigure{\label{2a}} 
	\subfigure{\label{2b}}
	\subfigure{\label{2c}}
	\subfigure{\label{2d}} 
	\subfigure{\label{2e}}
	\subfigure{\label{2f}} 
	\subfigure{\label{2g}} 
	\caption{(a) Schematic layout of the intense laser sources driven by two synchronized high-power Ti:sapphire amplifiers, each coupled with a dual-output optical parametric amplifier, following the initial proposal~\delaycite{Dong2018CPB}. The distinct nonlinear frequency conversion pathways and their corresponding spectral coverages are indicated. (b)–(g) Typical time-domain waveforms and corresponding Fourier-transformed spectra for the three high-field sources: (b) and (c) low-frequency THz pulses from optical rectification in LiNbO$_3$ crystal; (d) and (e) broadband THz pulses from the organic crystal DSTMS; (f) and (g) carrier-envelope-phase-stable MIR transients at a 10-$\mu$m central wavelength generated via difference-frequency-generation in GaSe crystal.} \label{2}
\end{figure*}

In THz-TDS measurements, the customized sample holder permits the simultaneous loading of multiple samples and references. The time-domain waveforms of the THz electric fields transmitted through the sample $E_{\mathrm{sam}}(t)$ and the reference $E_{\mathrm{ref}}(t)$ are recorded sequentially under identical experimental conditions. Following Fourier transformation, the complex transmission spectrum $\widetilde{T}(\omega)=\widetilde{E}_{\mathrm{sam}}(\omega)/\widetilde{E}_{\mathrm{ref}}(\omega)$ is obtained~\cite{Neu2018}. Notably, utilizing empty apertures with a 2.5-mm diameter, the platform demonstrates a near-unity transmission baseline with amplitude fluctuations less than 0.5\% across an effective spectral range of 0.1 to 3.5~THz (Fig. \ref{1c}). By iteratively solving the Fresnel equations that account for the sample thickness and internal multiple reflections, the complex optical constants are rigorously extracted without resorting to Kramers-Kronig transformations~\cite{Duvillaret1996JSTQE}.

\begin{table*}[t]
	\caption{Summary of key experimental parameters for the intense THz and MIR sources. Abbreviations: TPF (tilted-pulse-front), OR (optical rectification), DFG (difference-frequency generation). Symbols used: $\lambda$ (pump wavelength), $E_p$ (pump pulse energy), $\mathcal{E}$ (THz/MIR output pulse energy), $d$ (focal spot diameter), and $E_{\text{THz}}$ (peak electric field). The values represent typical performance parameters characterized under optimal focusing conditions. As a representative MIR example, the intense 12.5~$\mu$m pulse is generated via DFG between two signal beams (1380~nm and 1550~nm), with its corresponding driving and output parameters tabulated.}
	\label{tab1}
	\centering
	\renewcommand{\arraystretch}{1.3} 
	\begin{tabular*}{\textwidth}{@{\extracolsep{\fill}}lcccccccc}
		\toprule
		\multirow{2}{*}{\textbf{Source}} & 
		\multirow{2}{*}{\textbf{Freq.} (THz)} & 
		\multirow{2}{*}{\textbf{\makecell{Generation \\ Scheme}}} & 
		\multicolumn{2}{c}{\textbf{Pump Parameters}} & 
		\multirow{2}{*}{\textbf{Efficiency} (\%)} & 
		\multicolumn{3}{c}{\textbf{THz/MIR Output}} \\
		\cmidrule(lr){4-5} \cmidrule(lr){7-9}
		& & & {$\lambda$ (nm)} & {$E_p$ (mJ)} & & {$\mathcal{E}$ ($\mu$J)} & {$d$ ($\mu$m)} & {$E_{\text{THz}}$ (MV/cm)} \\
		\midrule
		LiNbO$_3$ & 0.1 -- 3 & TPF & 800 & 5.5 & 0.4 & 20 & 500 & 1 \\
		\addlinespace
		DSTMS & 1 -- 6 & OR & 1350 & 0.40 & 0.1 & 5 & 90 & Several \\
		\addlinespace
		GaSe & \makecell{20 -- 60 \\ (e.g., 24 THz / 12.5 $\mu$m)} & DFG & \makecell{1380\\1550} & \makecell{0.38\\0.30} & 1.5 & 10 & $<50$ & Tens \\
		\bottomrule
	\end{tabular*}
\end{table*}

\begin{figure*}[htb]
	\includegraphics[width=0.85\linewidth]{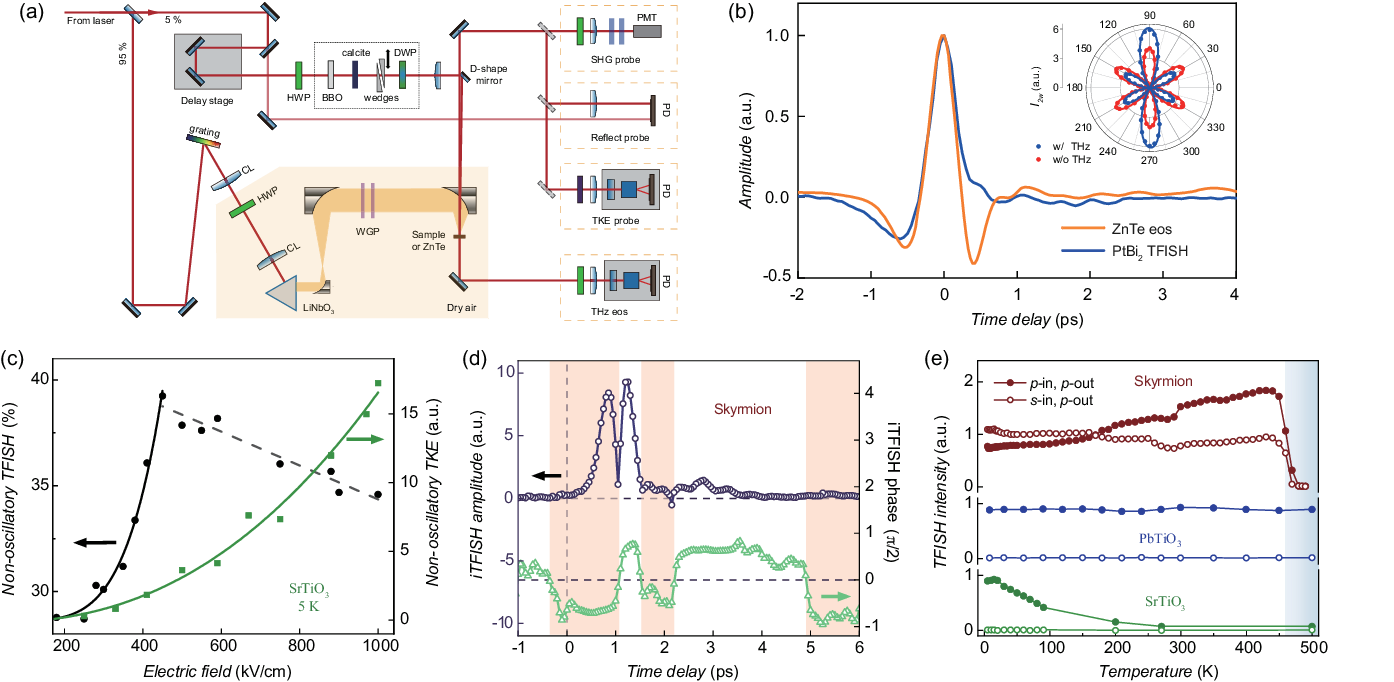}
	\centering		\vspace{-0.3cm}
	\subfigure{\label{3a}} 
	\subfigure{\label{3b}}
	\subfigure{\label{3c}}
	\subfigure{\label{3d}} 
	\subfigure{\label{3e}}
	\caption{
		(a) Schematic diagram of the THz-pump optical-probe setup incorporating an interferometric second-harmonic generation module for phase-sensitive detection. 
		(b) Transient SHG enhancement in the topological semimetal PtBi$_2$ following the incident THz waveform, with the inset displaying the SHG polar plot under a parallel configuration. Adapted with permission from Ref.~\delaycite{Gao2024PtBi2}. Copyrigh Optical Society of America. 
		(c) Field-dependent amplitudes of the non-oscillatory TFISH component and the TKE signal in SrTiO$_3$ at 5~K. Adapted from Ref.~\delaycite{SrTiO3Arxiv2024}. 
		(d) Time-domain iTFISH measurements of polar skyrmions in PbTiO$_3$/SrTiO$_3$ superlattices, resolving distinct phase inversions. 
		(e) Temperature dependence of the peak TFISH responses for polar skyrmions, alongside reference measurements on PbTiO$_3$ thin films and SrTiO$_3$ single crystals, measured up to 500~K. Panels (d) and (e) are adapted with permission from Ref.~\delaycite{Li2025Skyrmions}. Copyright 2025 Springer Nature.	} 	\label{3}
\end{figure*}

The obtained optical constants provide unambiguous signatures of charge and spin dynamics. For superconductors, the energy gap can be identified from the suppression in the real part of the optical conductivity below the transition temperature, as recently demonstrated in the disordered superconductor ZrN$_x$ film~\cite{Chen2023ZrN}. The derived superconducting gap follows the standard Bardeen-Cooper-Schrieffer (BCS) temperature dependence (Fig. \ref{1d}). This result indicates a conventional electron-phonon coupling mechanism, providing critical insights into the microscopic origin of its anomalous superconducting dome. For collective magnetic excitations, the application of high magnetic fields at low temperatures allows the probing of field-induced spin responses, which manifest as well-defined resonances in the complex magnetic susceptibility. A representative study focuses on the Kitaev quantum spin liquid candidate Na$_2$Co$_2$TeO$_6$~\cite{Shi2025Na2Co2TeO6}. As illustrated in the color map of Fig. \ref{1e}, under an in-plane magnetic field of 10~T, a broad fractionalized spin excitation continuum is clearly resolved. Remarkably, the excitation continuum persists up to 40~K, substantiating the survival of strong spin correlations and intense quantum fluctuations deep into the paramagnetic regime. Moreover, this platform facilitates the precise manipulation of intricate exchange interactions in rare-earth orthoferrites. Related studies include the discovery of anomalous splitting, merging, and re-splitting magnon behaviors in Gd$_{0.3}$Ho$_{0.7}$FeO$_3$~\cite{Li2026GdHoFeO3}, the magnetic-field control of electromagnons in Dy$_{0.9}$Nd$_{0.1}$FeO$_3$~\cite{Fu2024DyNdFeO3}, and the realization of strong THz cavity magnon-polaritons in Gd$_{0.5}$Ho$_{0.5}$FeO$_3$, where the high-quality crystal itself acts as a Fabry-P\'erot cavity~\cite{Chen2025GdHoFeO3}. As depicted in the contour plot of Fig. \ref{1f}, by dynamically tuning the magnon modes across the cavity resonances via external magnetic fields, anti-crossing dispersions with pronounced vacuum Rabi splitting are observed, confirming the strong light-matter interaction regime~\cite{Chen2025GdHoFeO3}. These compelling demonstrations validate the platform's exceptional capability to uncover the intrinsic ground states and cooperative excitations of quantum materials.

\subsection{Ultrafast Terahertz Nonlinear Spectroscopy}

\subsubsection{High-Intensity Terahertz and Mid-Infrared Sources}

To drive quantum materials far from equilibrium and enable the coherent control of their emergent properties, it is imperative to develop high-intensity THz sources~\cite{Nicoletti2016AOP,Salen2019PhysRep,Huang2026NRP}. The THz experimental unit (A4-2) is equipped with two synchronized Ti:sapphire amplifiers delivering 35-fs, 7-mJ pulses at a 1-kHz repetition rate, each coupled with a dual-output optical parametric amplifiers (OPAs). Through nonlinear frequency conversion, three intense THz sources have been established, as outlined in Fig. \ref{2}, spanning a broad spectral range from 0.1 to 60~THz.

First, for the low-frequency regime from 0.1 to 3~THz, vertically polarized THz pulses are generated in a MgO-doped LiNbO$_3$ crystal, which is pumped by the primary fundamental output of the first amplifier. A reflective diffraction grating is utilized to tilt the pump-pulse front to effectively compensate for the large velocity mismatch between the near-infrared pump and the generated THz wave~\cite{Hebling2008JOSAB}. The optimized configuration achieves a pump-to-THz energy conversion efficiency of approximately 0.4\%, providing a maximum THz pulse energy exceeding 20~$\mu$J~\cite{Wang2024RSI,Wang2026RSI}. We can tightly focus the THz beam to a 1/e$^2$ spot diameter of 0.5~mm, reaching a peak electric field larger than 1~MV/cm with a center frequency of 0.8~THz~\cite{peng2023terahertz}.

Second, to access higher THz frequencies from 1 to 6~THz, we employ optical rectification in organic crystals. Specifically, DSTMS possesses a large nonlinear optical coefficient but requires near-infrared driving pulses in the wavelength range of 1300 to 1500~nm to satisfy the phase-matching condition~\cite{Mansourzadeh2023OME}. Driven by the 400-$\mu$J, 1350-nm signal output from one of the optical parametric amplifiers (OPAs, TOPAS twins), the source generates broadband THz pulses with an initial measured energy of 5 $\mu$J. This value is obtained after the generation crystal through three 20-THz low-pass filters, which are necessary to block the residual pump. Since the shorter THz wavelengths enable a tighter focus, the THz beam is focused to a spot diameter of 90~$\mu$m. The resulting peak electric field reaches several MV/cm, offering exceptionally high field strengths for broadband nonlinear excitations~\cite{Shalaby2015NatCommun}.

Third, for the mid-infrared regime, pulses are generated via difference frequency generation (DFG) in GaSe crystal~\cite{Sell2008OL}. By adjusting the wavelengths of the two signal outputs from the TOPAS twins, the resulting MIR radiation can be continuously tuned from 5 to 15 $\mu$m (20 to 60~THz). For example, when pumped by the 1380-nm (380~$\mu$J) and 1550-nm (300~$\mu$J) signal beams, the DFG in 1-mm-thick GaSe crystal delivers a typical single-pulse energy of 10~$\mu$J at 12.5~$\mu$m. Because these pump beams originate from the twin OPAs sharing the same white-light seed, the MIR transients are passively carrier-envelope-phase stable~\cite{Liu2017OL}. The rapidly oscillating MIR electric fields can be resolved in the time domain via electro-optic sampling with sub-20-fs gating pulses. Benefiting from the substantially shorter wavelengths in the MIR regime, the pulses can be tightly focused to a spot size of $< 50~\mu$m. Consequently, the peak electric field reaches tens of MV/cm~\cite{Nicoletti2016AOP}. Coupled with its broad spectral tunability, this extreme field goes far beyond driving conventional nonlinear lattice dynamics. It provides a versatile tool for mode-selective resonant excitation, pushing specific phonon modes into highly anharmonic regimes and triggering the emergence of novel non-equilibrium states~\cite{Nicoletti2016AOP}.

The dual-amplifier laser system provides intense THz and MIR sources, with their optimal parameters achieved under free-space focusing conditions, as summarized in Table \ref{tab1}. The peak electric field of the THz and MIR pulses scales as $E_{\mathrm{THz}} \propto \sqrt{\mathcal{E}}/d$, where $\mathcal{E}$ represents the pulse energy and $d$ denotes the focal spot diameter~\cite{Reid05}. Integrating these high-intensity sources with extreme sample environments, such as low temperatures and high magnetic fields, inevitably attenuates the available field at the sample position. The reduction stems primarily from propagation losses through various optical elements and windows along the beam path, combined with the strict geometric constraints of cryostats and magnets that necessitate long-focal-length off-axis parabolic mirrors for focusing~\cite{Wang2024RSI}. To demonstrate the platform's advanced capabilities for nonlinear optical manipulation, representative probing techniques, including intense THz-pump optical-probe spectroscopy, THz high-harmonic generation, and THz two-dimensional coherent spectroscopy, are detailed in the following subsections.

\begin{figure*}
	\includegraphics[width=0.85\linewidth]{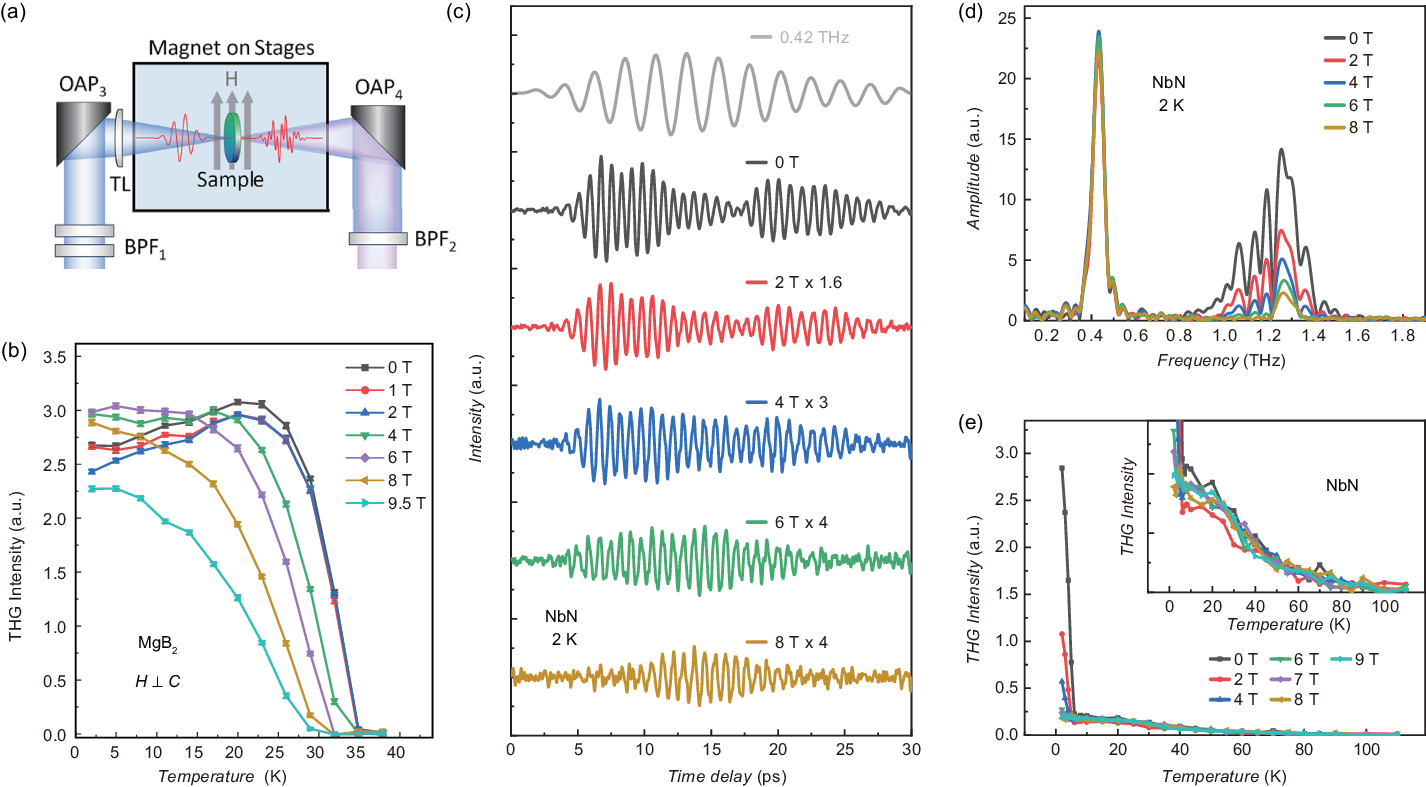}
	\centering		\vspace{-0.2cm}
	\subfigure{\label{4a}} 
	\subfigure{\label{4b}}
	\subfigure{\label{4c}}
	\subfigure{\label{4d}} 
	\subfigure{\label{4e}}
	\subfigure{\label{4f}} 
	\caption{
		(a) Detailed diagram of the tight-focusing geometry combining an off-axis parabolic (OAP) mirror and a high-density polyethylene lens (TL). Bandpass filters (BPFs) are used for multi-cycle pumping and THG extraction. 
		(b) Temperature dependence of the THG intensity in MgB$_2$ under various out-of-plane magnetic fields. Adapted with permission from Ref.~\delaycite{Wang2024RSI}. Copyright 2024 AIP Publishing. 
		(c) Time-domain THG waveforms and (d) the corresponding Fourier-transformed spectra for a disordered NbN thin film measured at 2~K under various magnetic fields. 
		(e) Temperature and magnetic-field evolution of the THG intensity in the disordered NbN film, with the inset providing a magnified view of the magnetic-field dependence in the high-temperature normal state. Panels (c)–(e) are adapted with permission from Ref.~\delaycite{Wang2025NbN}. Copyright 2026 by the American Physical Society.
	} 
	\label{4}
\end{figure*}

\subsubsection{Terahertz-Pump Optical-Probe Spectroscopy}
In a typical pump-probe geometry (Fig. \ref{3a}), an intense THz pump and a temporally delayed optical probe are focused collinearly onto the sample at near-normal incidence to trace the ultrafast nonequilibrium dynamics. We can selectively monitor distinct transient optical responses. Specifically, transient reflectivity ($\Delta$R/R) tracks quasiparticle relaxation, the THz Kerr effect (TKE) measures transient birefringence; and THz-field-induced second harmonic generation (TFISH) probes inversion symmetry breaking. Because conventional TFISH measures only the intensity of the second-harmonic light, the phase information of the emitted field is inherently lost. To overcome this limitation, an interferometric TFISH (iTFISH) module is implemented. A $\beta$-barium borate (BBO) crystal generates a 400 nm local oscillator (LO), while a calcite crystal compensates the bulk group velocity dispersion between the 800 nm probe and the 400 nm LO, with a pair of fused silica wedges providing precise phase modulation. The interference between the TFISH from the sample and the LO enables phase sensitive detection, thereby extracting the sign of specific nonlinear tensor elements~\cite{Lin2024ACSPho, Li2025Skyrmions}.

Combining intense THz excitations with these optical probes facilitates the direct manipulation and detection of hidden quantum phases. In the topological semimetal PtBi$_2$, intense THz fields perturb the linear dispersion bands near the Fermi level, producing a 45\% enhancement in the transient SHG signal that follows the THz waveform (Fig. \ref{3b})~\cite{Gao2024PtBi2}. In the quantum paraelectric SrTiO$_3$, resonant excitation of the ferroelectric soft mode initiates complex structural dynamics. While previous studies suggested a THz-induced ferroelectric state~\cite{Li2019Science}, driving the soft mode into a second-order excited state with elevated THz fields exceeding 500~kV/cm unveils a reentrant phase transition into a hidden quantum paraelectric phase, as identified by the contrasting field-dependent behaviors of the TFISH and TKE signals (Fig. \ref{3c})~\cite{SrTiO3Arxiv2024}. Furthermore, THz-driven coherent control extends to topological polar skyrmions in PbTiO$_3$/SrTiO$_3$ superlattices, where the THz pulses couple to the collective modes of the skyrmion walls, giving rise to a hidden phase with a transient macroscopic planar polarization~\cite{Li2025Skyrmions}. Crucially, the iTFISH technique successfully resolves distinct phase inversions in the time domain (Fig. \ref{3d}), providing direct evidence of the polarization flipping governed by the competition among multiple excited collective modes. The light-induced phase transition is sustained over a broad temperature range up to 470~K (Fig. \ref{3e}), due to the intrinsic topological protection of the skyrmion collective modes rather than conventional soft mode dynamics~\cite{Li2025Skyrmions}. The precise extraction of these diverse and subtle transient phenomena, ranging from perturbative electronic responses to complex collective mode competitions, demonstrates the capability of the TPOP platform to capture ultrafast quantum dynamics across broad temperature ranges under strong driving fields ($\sim$1 MV/cm).

\begin{figure*}
	\includegraphics[width=0.85\linewidth]{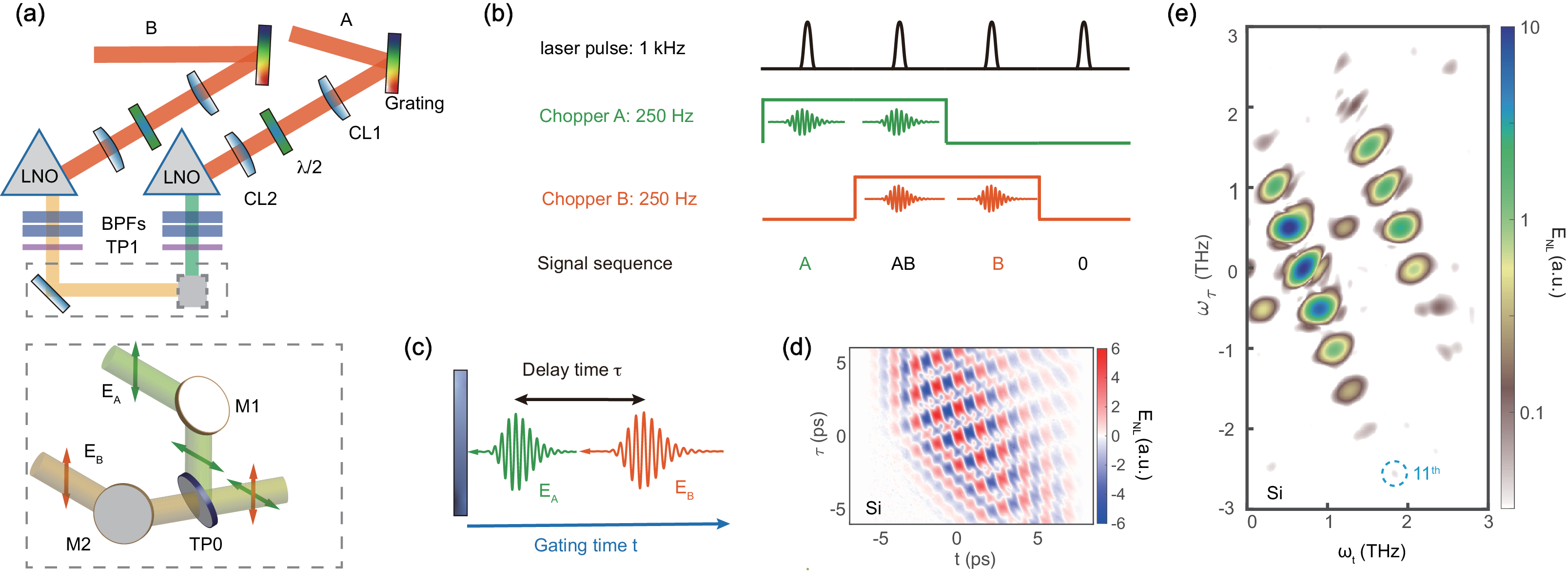}
	\centering		\vspace{-0.2cm}
	\subfigure{\label{5a}} 
	\subfigure{\label{5b}}
	\subfigure{\label{5c}}
	\subfigure{\label{5d}} 
	\subfigure{\label{5e}}
	\subfigure{\label{5f}} 
	\caption{(a) Experimental design of the 2D-THz coherent spectroscopy platform employing dual LiNbO$_3$ crystals and a customized periscope-based beam combiner. (b) Pulse sequence and modulation scheme for isolating the pure nonlinear signal in 2D-THz experiments. (c) Timing sequence of the two-pulse excitation defining the time delays. (d) Measured 2D time-domain and (e) corresponding frequency-domain nonlinear responses of n-doped silicon under 0.5-THz and 0.7-THz dual-frequency excitations at room temperature. Adapted with permission from Ref.~\delaycite{Wang2026RSI}. Copyright 2024 AIP Publishing.} \label{5}
\end{figure*}

\subsubsection{Terahertz High-Harmonic Generation Spectroscopy}
Nonlinear THz high-harmonic generation spectroscopy requires intense narrowband excitations to resonantly drive nonperturbative electron dynamics and collective excitations~\cite{Salen2019PhysRep, Yang2023NatRevMater}. While intense narrowband THz radiation is predominantly available at free-electron laser facilities which have enabled THz high harmonic-generation in quantum materials~\cite{Hafez2018Nature, Chu2023NatCommun}, tabletop systems can utilize specific bandpass filters to extract the desired frequency components from a broadband source, at the cost of attenuating the electric field strength. Integrating these tailored excitations with low temperatures and high magnetic fields provides an essential means to disentangle competing quantum orders. However, the large physical dimensions of the superconducting magnet cryostat enforce long-focal-length optics, resulting in a large diffraction-limited spot size that further weakens the THz electric field.

To address this challenge, we employ a hybrid focusing geometry comprising an off-axis parabolic mirror and a high-density polyethylene lens to tightly focus the high-energy THz pulses generated from the LiNbO$_3$ crystal into the magnet center (Fig. \ref{4a})~\cite{Wang2024RSI}. At the sample position, the peak driving electric field reaches 500~kV/cm for monocycle waveforms and 30~kV/cm for 0.5-THz narrowband excitations. To demonstrate the capabilities of the magneto-THz nonlinear platform, we investigated the THG response of the two-band superconductor MgB$_2$~\cite{Wang2024RSI}. Utilizing the motorized rotation stage (Fig. \ref{1b}), \textit{in situ} azimuthal rotation was performed to resolve the polarization dependence of the nonlinear emission without altering the incident THz polarization. Furthermore, applying an out-of-plane magnetic field rapidly suppresses the smaller $\pi$-band gap, effectively quenching the THG resonance peak (Fig. \ref{4b}). By integrating strong-field THz capabilities with extreme sample environments (up to 10 T and down to 1.5 K), this instrument overcomes the focal distance limitations of conventional table-top systems, providing a robust platform for exploring magnetic-field-dependent nonlinear quantum dynamics.

The capability of this platform to resolve diverse higher-order nonlinear optical phenomena is exemplified by the study of disordered NbN thin films near the superconductor-insulator transition~\cite{Wang2025NbN}. Upon excitation with 0.42-THz multicycle pulses, the sample emits a third-harmonic signal that is clearly resolved in both the time and frequency domains (Fig. \ref{4c} and \ref{4d}). Strikingly, an anomalous THG signal persists above the superconducting transition temperature of approximately 6~K and remains unchanged even when a 9-T magnetic field fully suppresses the global phase coherence (Fig. \ref{4e}). Such magnetic-field insensitivity demonstrates that the normal-state nonlinearity originates from disorder-induced band structure modifications rather than superconducting fluctuations. Below the critical temperature, the THG emission from the driven Higgs mode interferes with the coexisting normal-state channel, manifesting as a temporal beating waveform and a broadened multi-peak spectrum~\cite{Wang2025NbN}.

\begin{figure*}
	\includegraphics[width=0.8\linewidth]{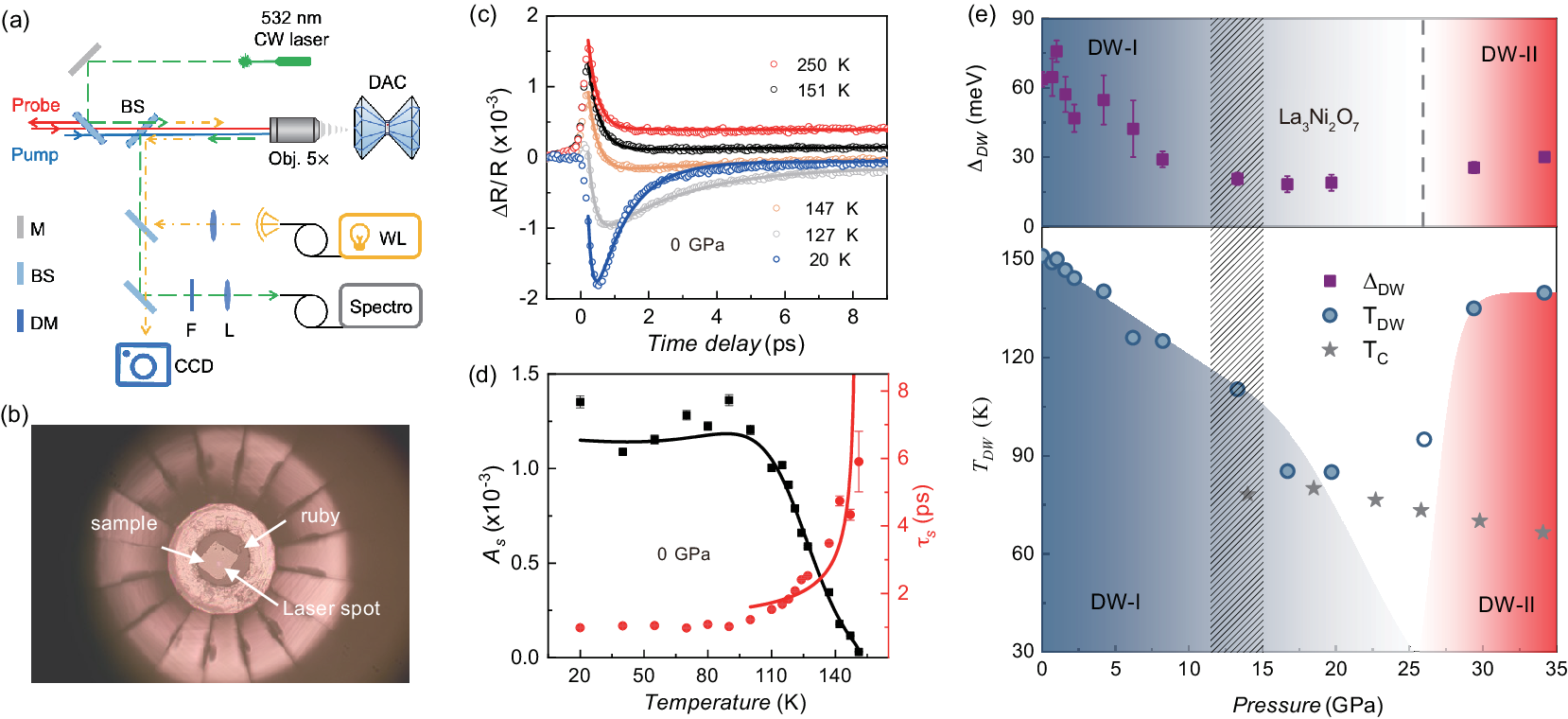}
	\centering		\vspace{-0.2cm}
	\subfigure{\label{6a}} 
	\subfigure{\label{6b}}
	\subfigure{\label{6c}}
	\subfigure{\label{6d}} 
	\subfigure{\label{6e}}
	\subfigure{\label{6f}} 
	\caption{(a) Instrumental layout of the customized optical pump-probe setup featuring a six-beam coaxial integration. Adapted from Ref.~\delaycite{Meng2026Ultrafast}. 
		(b) Photograph of the sample and ruby ball loaded inside a diamond-anvil cell. 
		(c) Temperature-dependent transient reflectivity profiles of La$_3$Ni$_2$O$_7$ measured near ambient pressure. 
		(d) Extracted relaxation amplitude and decay time together with the Rothwarf-Taylor model fit. 
		(e) Temperature-pressure phase diagram illustrating the evolution of the density-wave orders in pressurized La$_3$Ni$_2$O$_7$. Panels (c)–(e) are adapted from Ref.~\delaycite{Meng2024NatCommun}, under the terms of the Creative Commons CC BY license.	} 	\label{6}
\end{figure*}

\subsubsection{Terahertz Two-Dimensional coherent Spectroscopy}
Extending beyond conventional one-dimensional nonlinear techniques, two-dimensional THz (2D-THz) coherent spectroscopy provides a transformative method for probing and controlling multicorrelations in quantum matter~\cite{Huang2026NRP}. It measures the nonlinear response arising from the interaction between two intense THz pulses as a function of both the inter-pulse delay and the detection time (Fig. \ref{5b} and \ref{5c}). Applying a two-dimensional Fourier transform then converts the temporal signals into the frequency domain, which can disentangle multi-order quantum pathways and isolate overlapping collective modes. To realize these demanding measurements, we have developed a versatile 2D-THz platform equipped with dual high-intensity sources generated from separate LiNbO$_3$ crystals~\cite{Wang2026RSI}. A customized periscope-based beam combiner merges the high-energy THz beams, ensuring spatiotemporal overlap while preserving their peak electric fields (Fig. \ref{5a}). The dual-source design provides the experimental flexibility to independently control the polarization, spectrum, and amplitude of each driving pulse. To demonstrate the instrument's performance, we investigated the nonlinear response of silicon under dual-frequency excitations at 0.5~THz and 0.7~THz. The 2D time-domain signal emerges exclusively during the temporal overlap of both pump pulses (Fig. \ref{5c}), confirming its nonlinear origin. The resulting frequency map (Fig. \ref{5d}) clearly reveals various complex wave-mixing processes up to the eleventh-order at room temperature~\cite{Wang2026RSI}. While conventional one-dimensional spectroscopy typically requires the extreme field strengths of free-electron lasers and cryogenic cooling to detect comparable high-order harmonics, this tabletop 2D configuration effectively isolates overlapping wave-mixing pathways to resolve these distinct nonlinearities directly. Furthermore, integrating this dual-source architecture with the 10-T superconducting magnet enables magneto-2D-THz coherent spectroscopy, offering an unprecedented opportunity to explore many-body dynamics and field-tuned collective excitations, such as magnons, fractionalized spins, and Landau-Dirac fermions in complex quantum materials~\cite{Huang2026NRP}.

\subsection{Ultrafast Optical Pump-Probe Spectroscopy under High Pressures and Low Temperatures}
While THz pulses face diffraction limits restricting their focal spots to the millimeter scale, near-infrared and visible beams can be tightly focused to microscopic dimensions, enabling integration with high-pressure techniques based on diamond-anvil cells. To explore the ultrafast dynamics under high pressures, we have developed a customized optical pump-probe setup operating in a collinear back-reflection geometry~\cite{Yang2024PRB}. As illustrated in Fig. \ref{6a}, the optical system coaxially combines six distinct beams utilizing a rigid cage system\cite{Meng2026Ultrafast}. Driven by a 50-kHz Yb-based laser system, the time-resolved reflectivity module employs a non-degenerate 400-nm pump and 800-nm probe. These ultrafast pulses are integrated with a white-light path for sample imaging  and a 532-nm continuous-wave laser coupled to a high-resolution spectrometer for ruby fluorescence pressure calibration (Fig. \ref{6b}). By focusing all incident beams and collecting the reflected signals through a single long-working-distance objective lens, the configuration enables ultrafast micro spectroscopy at pressures up to 35~GPa and low temperatures down to 10~K.

Utilizing this high-pressure platform, the complex density-wave orders in the bilayer nickelate superconductor La$_3$Ni$_2$O$_7$ were systematically investigated~\cite{Meng2024NatCommun}. At low pressures, the temperature-dependent transient reflectivity exhibits a distinct phonon bottleneck effect upon cooling, characterized by the emergence of a slow relaxation component with a negative amplitude (Fig. \ref{6c}). By applying the Rothwarf-Taylor model to the extracted relaxation amplitude and decay time (Fig. \ref{6d}), a density-wave-like gap of approximately 66 meV is determined. Upon compression, the initial order is progressively suppressed up to 13.3~GPa, and the quasi-divergent relaxation behavior becomes weak before vanishing completely near 26~GPa. Above~29.4 GPa, the phonon bottleneck effect reemerges at a transition temperature of approximately 135~K, indicating the formation of a distinct charge-density-wave phase. The resulting phase diagram (Fig. \ref{6e}) illustrates the high-pressure evolution of the density-wave orders, providing critical experimental insights into their underlying correlation with superconductivity in pressurized nickelates~\cite{Meng2024NatCommun}. The application of the OPOP platform was further extended to the cuprate superconductor Bi$_2$Sr$_2$CaCu$_2$O$_{8+\delta}$ up to 37 GPa~\cite{Meng2026Ultrafast}. By resolving the distinct quasiparticle relaxation channels at low temperatures, the recent study successfully decoupled the intertwined pseudogap and superconducting phases across an extended high-pressure phase diagram, establishing the platform's capability to probe complex many-body dynamics under extreme conditions.

\section{Conclusion and Perspectives}

Over the past years, the THz experimental unit (A4-2) at SECUF has successfully transitioned from an initial conceptual design into a comprehensive research platform~\cite{Dong2018CPB}. By coupling ultrafast optical technologies with extreme sample environments, the facility provides a versatile suite of spectroscopic capabilities, currently primarily utilizing LiNbO$_3$-based intense THz sources for various pump-probe configurations. The instrumental developments across these setups include achieving peak THz fields of $\sim$1 MV/cm in the TPOP system, integrating 10-T magnetic fields down to 1.5~K for magneto-THz-THG measurements, realizing 2D-THz coherent spectroscopy driven by dual independently controlled intense pulses, and incorporating high-pressure and low-temperature environments into the OPOP configuration. As demonstrated by recent experimental studies, these established capabilities provide a rigorous and highly competitive experimental foundation for exploring ultrafast quantum dynamics.

To further expand the spectral coverage, polarization states, and excitation capabilities, continuous efforts are dedicated to developing advanced light sources. First, various nonlinear optical materials, including organic crystals and SiC, are being implemented to generate highly intense transients, effectively bridging the conventional frequency gap between 5 and 15 THz~\cite{Liu2017OL,Fischer17OL}. Furthermore, advanced nonlinear optical techniques, employing chirped-pulse difference frequency generation, are being developed to generate narrowband THz and MIR pulses~\cite{VicarioAPL20,Cartella17}. Additionally, as existing sources are predominantly linearly polarized, dedicated efforts are underway to provide intense circularly polarized THz and MIR fields, offering a powerful tool for manipulating chiral and magnetic excitations\cite{Basini2024Nature,Minakova2026NatPhys}. Finally, integrating these versatile light sources with extreme sample environments encompassing temperatures down to 1.5 K, magnetic fields up to 10 T, and quasi-hydrostatic pressures up to 35 GPa constitutes a powerful framework for exploring non-equilibrium dynamics.

As a user facility, the THz experimental unit actively seeks broad scientific collaborations. We sincerely invite researchers from the global scientific community to submit experimental proposals and utilize this integrated platform to coherently manipulate complex systems and unravel the underlying physics of emergent phenomena across diverse scientific disciplines.

\section{Acknowledgements}
This work was supported by the National Key Research and Development Program of China (Grants No. 2024YFA1611300) and the National Natural Science Foundation of China (Grants No. 12574349). We gratefully acknowledge the user community for their collaborative efforts at the THz experimental unit, whose contributions have led to both the representative results discussed in this review and many other ongoing research projects. This work was carried out at the Synergetic Extreme Condition User Facility (SECUF, https://cstr.cn/31123.02.SECUF).

\bibliography{THz_CPB_WOS3.bib}

@article{Leitenstorfer2023,
	title="{The 2023 terahertz science and technology roadmap}",
	Author = {Leitenstorfer, Alfred and Moskalenko, Andrey S. and Kampfrath, Tobias and Kono, Junichiro and Castro-Camus, Enrique and Peng, Kun and Qureshi, Naser and Turchinovich, Dmitry and Tanaka, Koichiro and Markelz, Andrea G. and Havenith, Martina and Hough, Cameron and Joyce, Hannah J. and Padilla, Willie J. and Zhou, Binbin and Kim, Ki-Yong and Zhang, Xi-Cheng and Jepsen, Peter Uhd and Dhillon, Sukhdeep and Vitiello, Miriam and Linfield, Edmund and Davies, A. Giles and Hoffmann, Matthias C. and Lewis, Roger and Tonouchi, Masayoshi and Klarskov, Pernille and Seifert, Tom S. and Gerasimenko, Yaroslav A. and Mihailovic, Dragan and Huber, Rupert and Boland, Jessica L. and Mitrofanov, Oleg and Dean, Paul and Ellison, Brian N. and Huggard, Peter G. and Rea, Simon P. and Walker, Christopher and Leisawitz, David T. and Gao, Jian Rong and Li, Chong and Chen, Qin and Valu\v{s}is, Gintaras and Wallace, Vincent P. and Pickwell-MacPherson, Emma and Shang, Xiaobang and Hesler, Jeffrey and Ridler, Nick and Renaud, Cyril C. and Kallfass, Ingmar and Nagatsuma, Tadao and Zeitler, J. Axel and Arnone, Don and Johnston, Michael B. and Cunningham, John},
	journal={J. Phys. D: Appl. Phys.},
	volume={56},
	number={22},
	pages={223001},
	year={2023},
	publisher={IOP Publishing},
	url={https://doi.org/10.1088/1361-6463/acbe4c}
}

@article{Basov2011,
title = {Electrodynamics of correlated electron materials},
author = {Basov, D. N. and Averitt, Richard D. and van der Marel, Dirk and Dressel, Martin and Haule, Kristjan},
journal = {Rev. Mod. Phys.},
volume = {83},
issue = {2},
pages = {471--541},
numpages = {0},
year = {2011},
month = {Jun},
publisher = {American Physical Society},
doi = {10.1103/RevModPhys.83.471},
url = {https://link.aps.org/doi/10.1103/RevModPhys.83.471}
}

@article{UlbrichtRevModPhys2011,
	title = {Carrier dynamics in semiconductors studied with time-resolved terahertz spectroscopy},
	author = {Ulbricht, Ronald and Hendry, Euan and Shan, Jie and Heinz, Tony F. and Bonn, Mischa},
	journal = {Rev. Mod. Phys.},
	volume = {83},
	issue = {2},
	pages = {543--586},
	numpages = {0},
	year = {2011},
	month = {Jun},
	publisher = {American Physical Society},
	doi = {10.1103/RevModPhys.83.543},
	url = {https://link.aps.org/doi/10.1103/RevModPhys.83.543}
}

@article{Basini2024Nature,
	author    = {Basini, M. and Pancaldi, M. and Wehinger, B. and Udina, M. and  Unikandanunni, V. and Tadano, T. and Hoffmann, M. C. and Balatsky, A. V. and Bonetti, S.},
	title     = {Terahertz electric-field-driven dynamical multiferroicity in {SrTiO}$_3$},
	journal   = {Nature},
	volume    = {628},
	number    = {8008},
	pages     = {534--539},
	year      = {2024},
	doi       = {10.1038/s41586-024-07175-9},
	url       = {https://doi.org/10.1038/s41586-024-07175-9}
}

@article{Minakova2026NatPhys,
	author    = {Minakova, Olga and Paiva, Carolina and Frenzel, Maximilian and Spencer, Michael S. and Urban, Joanna M. and Ringkamp, Christoph and Wolf, Martin and Mussler, Gregor and Juraschek, Dominik M. and Maehrlein, Sebastian F.},
	title     = {Observation of angular momentum transfer among crystal lattice modes},
	journal   = {Nat. Phys.},
	year      = {2026},
	month     = {May},
	day       = {12},
	url       = {https://doi.org/10.1038/s41567-026-03274-8}
}

@article{VicarioAPL20,
 	author = {Vicario, C. and Trisorio, A. and Allenspach, S. and R\"{u}egg, C. and Giorgianni, F.},
 	title = {Narrow-band and tunable intense terahertz pulses for mode-selective coherent phonon excitation},
 	journal = {Appl. Phys. Lett.},
 	volume = {117},
 	number = {10},
 	pages = {101101},
 	year = {2020},
 	month = {09},
 	issn = {0003-6951},
 	doi = {10.1063/5.0015612},
 	url = {https://doi.org/10.1063/5.0015612},
 }

@article{Reid05,
	author = {Matthew Reid and Robert Fedosejevs},
	journal = {Appl. Opt.},
	number = {1},
	pages = {149--153},
	publisher = {Optica Publishing Group},
	title = {Quantitative comparison of terahertz emission from (100) {InAs} surfaces and a {GaAs} large-aperture photoconductive switch at high fluences},
	volume = {44},
	month = {Jan},
	year = {2005},
	url = {https://opg.optica.org/ao/abstract.cfm?URI=ao-44-1-149},
	doi = {10.1364/AO.44.000149},
}

@article{Cartella17,
	author = {A. Cartella and T. F. Nova and A. Oriana and G. Cerullo and M. F\"{o}rst and C. Manzoni and A. Cavalleri},
	journal = {Opt. Lett.},
	number = {4},
	pages = {663--666},
	publisher = {Optica Publishing Group},
	title = {Narrowband carrier-envelope phase stable mid-infrared pulses at wavelengths beyond 10 $\mu$m by chirped-pulse difference frequency generation},
	volume = {42},
	month = {Feb},
	year = {2017},
	url = {https://opg.optica.org/ol/abstract.cfm?URI=ol-42-4-663},
	doi = {10.1364/OL.42.000663},
}

@article{Fischer17OL,
	author = {Marco P. Fischer and Johannes B\"{u}hler and Gabriel Fitzky and Takayuki Kurihara and Stefan Eggert and Alfred Leitenstorfer and Daniele Brida},
	journal = {Opt. Lett.},
	number = {14},
	pages = {2687--2690},
	publisher = {Optica Publishing Group},
	title = {Coherent field transients below 15 {THz} from phase-matched difference frequency generation in {4H-SiC}},
	volume = {42},
	month = {Jul},
	year = {2017},
	url = {https://opg.optica.org/ol/abstract.cfm?URI=ol-42-14-2687},
	doi = {10.1364/OL.42.002687},
}

@article{kampfrath2013resonant,
	title={Resonant and nonresonant control over matter and light by intense terahertz transients},
	author={Kampfrath, Tobias and Tanaka, Koichiro and Nelson, Keith A.},
	journal={Nat. Photon.},
	volume={7},
	number={9},
	pages={680--690},
	year={2013},
	publisher={Nature Publishing Group},
	doi={10.1038/nphoton.2013.184},
	url={https://www.nature.com/articles/nphoton.2013.184}
}

@article{Neu2018,
	Author = {Neu, Jens and Schmuttenmaer, Charles A.},
	Title = "{Tutorial: An introduction to terahertz time domain spectroscopy (THz-TDS)}",
	Journal = {J. Appl. Phys.},
	Year = {2018},
	Volume = {124},
	Number = {23},
	Month = {DEC 21},
	pages = {231101},
	ISSN = {0021-8979},
	EISSN = {1089-7550},
	ResearcherID-Numbers = {Neu, Jens/S-5503-2018},
	ORCID-Numbers = {Neu, Jens/0000-0002-1054-0444},
	Unique-ID = {WOS:000454217800001},
	url={https://doi.org/10.1063/1.5047659}
}

@article{Yang2023NatRevMater,
	Author = {Yang, Chia-Jung and Li, Jingwen and Fiebig, Manfred and Pal, Shovon},
	title = "{Terahertz control of many-body dynamics in quantum materials}",
	journal = {Nat. Rev. Mater.},
	volume = {8},
	pages = {518--532},
	year = {2023},
	url = {https://doi.org/10.1038/s41578-023-00566-w}
}

@article{Salen2019PhysRep,
	author = {Sal{\'e}n, P. and Basini, M. and Bonetti, S. and Hebling, J. and Krasilnikov, M. and Nikitin, A. Y. and Shamuilov, G. and Tibai, Z. and Zhaunerchyk, V. and Goryashko, V.},
	title = "{Matter manipulation with extreme terahertz light: Progress in the enabling THz technology}",
	journal = {Phys. Rep.},
	volume = {836-837},
	pages = {1--74},
	year = {2019},
	url = {https://doi.org/10.1016/j.physrep.2019.09.002}
}

@article{peng2023terahertz,
	title={Terahertz-Field-Induced Second Harmonic Generation in Weyl Semimetal {TaAs}},
	author={Peng, P. and Li, Z. L. and Wang, X. B.},
	journal={Chin. J. Lasers},
	volume={50},
	number={17},
	pages={1714016},
	year={2023},
	publisher={Chinese Laser Press},
	doi={10.3788/CJL230830}
}

@article{Meng2026Ultrafast,
	title = {Ultrafast decoupling of the pseudogap from superconductivity in a pressurized cuprate},
	author = {Meng, Yanghao and Mao, Wenjin and Chen, Liucheng and Chia, Elbert E. M. and Yang, Yifeng and Luo, Jianlin and Zhao, Lin and Zhou, Xingjiang and Yu, Xiaohui and Wang, X. B.},
	year = {2026},
	Journal = {arXiv:2604.10207},
	url={https://arxiv.org/abs/2604.10207}, 
}

@article{Hafez2018Nature,
	author = {Hafez, H. A. and Kovalev, S. and Deinert, J.-C. and Mics, Z. and Green, B. and Awari, N. and Chen, M. and Germanskiy, S. and Lehnert, U. and Teichert, J. and Wang, Z. and Tielrooij, K.-J. and Liu, Z. and Chen, Z. and Narita, A. and M{\"u}llen, K. and Bonn, M. and Gensch, M. and Turchinovich, D.},
	title = "{Extremely efficient terahertz high-harmonic generation in graphene by hot Dirac fermions}",
	journal = {Nature},
	Year = {2018},
	Volume = {561},
	Number = {7724},
	Pages = {507},
	url = {https://doi.org/10.1038/s41586-018-0508-1}
}

@article{Chu2023NatCommun,
	Author = {Chu, Hao and Kim, Min-Jae and Katsumi, Kota and Kovalev, Sergey and Dawson, Robert David and Schwarz, Lukas and Yoshikawa, Naotaka and Kim, Gideok and Putzky, Daniel and Li, Zhi Zhong and Raffy, Helene and Germanskiy, Semyon and Deinert, Jan-Christoph and Awari, Nilesh and Ilyakov, Igor and Green, Bertram and Chen, Min and Bawatna, Mohammed and Cristiani, Georg and Logvenov, Gennady and Gallais, Yann and Boris, Alexander V. and Keimer, Bernhard and Schnyder, Andreas P. and Manske, Dirk and Gensch, Michael and Wang, Zhe and Shimano, Ryo and Kaiser, Stefan},
	Title = "{Phase-resolved Higgs response in superconducting cuprates}",
	Journal = {Nat. Commun.},
	Year = {2020},
	Volume = {11},
	Number = {1},
        Pages = {1793},
	Month = {APR 14},
	url = {https://doi.org/10.1038/s41467-020-15613-1},
	Article-Number = {1793},
	EISSN = {2041-1723}
}

@article{Wang2024RSI,
	Author = {Wang, X. B. and Wang, H. and Yuan, J. Y. and Zeng, X. Y. and Cheng, L. and Qi, J. and Luo, J. L. and Dong, T. and Wang, N. L.},
	Title = "{Table-top laser-based terahertz high harmonic generation spectroscopy under magnetic fields and low temperatures}",
	Journal = {Rev. Sci. Instrum.},
	Year = {2024},
	Volume = {95},
	Number = {10},
	Month = {OCT 1},
        Pages = {103007},
	url = {https://doi.org/10.1063/5.0215129},
	Article-Number = {103007},
	ISSN = {0034-6748},
	EISSN = {1089-7623},
	ResearcherID-Numbers = {Wang, Nan-Lin/HPG-1600-2023 Yuan, Jiayu/JQI-6524-2023},
	ORCID-Numbers = {Wang, Hao/0009-0001-7366-5121 Wang, Nan-Lin/0000-0003-2584-9265 Yuan, Jiayu/0000-0002-9849-5237 Cheng, Liang/0000-0002-3514-2207 Wang, Xinbo/0009-0001-0565-3911},
	Unique-ID = {WOS:001340448500005},
}

@article{Wang2025NbN,
	title = {Anomalous Terahertz Nonlinearity in Disordered $s$-Wave Superconductor Close to the Superconductor-Insulator Transition},
	author = {Wang, Hao and Yuan, Jiayu and Shi, Hongkai and Li, Haojie and Jia, Xiaoqing and Song, Xiaohui and Shi, Liyu and Wu, Tianyi and Yue, Li and Li, Yangmu and Jin, Kui and Wu, Dong and Luo, Jianlin and Wang, Xinbo and Dong, Tao and Wang, Nan-Lin},
	journal = {Phys. Rev. Lett.},
	volume = {136},
	issue = {14},
	pages = {146004},
	numpages = {7},
	year = {2026},
	month = {Apr},
	publisher = {American Physical Society},
	doi = {10.1103/t5wx-z86b},
	url = {https://link.aps.org/doi/10.1103/t5wx-z86b}
	}

@article{Wang2026RSI,
	Author = {Wang, X. B. and Shi, L. Y. and Zhang, S. J. and Yuan, J. Y. and Li, Y. T. and Luo, J. L. and Dong, T. and Wang, N. L.},
	Title = "{A versatile two-dimensional terahertz spectroscopy platform with dual independently controlled intense pulses}",
	Journal = {Rev. Sci. Instrum.},
	Year = {2026},
	Volume = {97},
	Number = {3},
        Pages = {033001},
	Month = {MAR 1},
	url = {https://doi.org/10.1063/5.0281877},
	Article-Number = {033001},
	ISSN = {0034-6748},
	EISSN = {1089-7623},
	Unique-ID = {WOS:001705433600001},
}

@article{Shalaby2015NatCommun,
	Author = {Shalaby, Mostafa and Hauri, Christoph P.},
	Title = "{Demonstration of a low-frequency three-dimensional terahertz bullet with extreme brightness}",
	Journal = {Nat. Commun.},
	Year = {2015},
	Volume = {6},
        Pages = {5976},
	Month = {JAN},
	url = {https://doi.org/10.1038/ncomms6976},
	Article-Number = {5976},
	ISSN = {2041-1723},
	ResearcherID-Numbers = {Hauri, Christoph/B-6192-2013},
	Unique-ID = {WOS:000348811700007},
}

@article{Mansourzadeh2023OME,
	Author = {Mansourzadeh, Samira and Vogel, Tim and Omar, Alan and Buchmann, Tobias O. and Kelleher, Edmund J. R. and Jepsen, Peter U. and Saraceno, Clara J.},
	Title = "{Towards intense ultra-broadband high repetition rate terahertz sources based on organic crystals {[}Invited]}",
	Journal = {Opt. Mater. Express},
	Year = {2023},
	Volume = {13},
	Number = {11},
	Pages = {3287-3308},
	Month = {NOV 1},
	url = {https://doi.org/10.1364/OME.502209},
	ISSN = {2159-3930},
	ResearcherID-Numbers = {Saraceno, Clara/K-4790-2016 Kelleher, Edmund/AAB-9589-2020 Jepsen, Peter Uhd/AAH-5439-2021},
	ORCID-Numbers = {Saraceno, Clara/0000-0002-7369-9057 Kelleher, Edmund/0000-0003-4219-4926 Mansourzadeh, Samira/0000-0002-2344-1525 Buchmann, Tobias/0000-0002-7007-3593 Jepsen, Peter Uhd/0000-0003-3915-1167 Vogel, Tim/0000-0002-0539-4331 Omar, Alan/0000-0003-0855-4641},
	Unique-ID = {WOS:001107260800008},
}

@article{Basov2017,
	Author = {Basov, D. N. and Averitt, R. D. and Hsieh, D.},
	Title = "{Towards properties on demand in quantum materials}",
	Journal = {Nat. Mater.},
	Year = {2017},
	Volume = {16},
	Number = {11},
	Pages = {1077-1088},
	Month = {NOV},
	url = {https://doi.org/10.1038/NMAT5017},
	ISSN = {1476-1122},
	EISSN = {1476-4660},
	ResearcherID-Numbers = {Averitt, Richard/AAX-4178-2021},
	ORCID-Numbers = {Hsieh, David/0000-0002-0812-955X Averitt, Richard/0000-0003-0451-1935},
	Unique-ID = {WOS:000413668800011},
}

@article{deLaTorre2021,
	Author = {de la Torre, Alberto and Kennes, Dante M. and Claassen, Martin and Gerber, Simon and McIver, James W. and Sentef, Michael A.},
	Title = "{Colloquium: Nonthermal pathways to ultrafast control in quantum materials}",
	Journal = {Rev. Mod. Phys.},
	Year = {2021},
	Volume = {93},
	Number = {4},
	Month = {OCT 14},
	url = {https://doi.org/10.1103/RevModPhys.93.041002},
	pages = {041002},
	ISSN = {0034-6861},
	EISSN = {1539-0756},
	ResearcherID-Numbers = {Kennes, Dante/AAJ-7286-2020 Gerber, Simon/A-4566-2012 Sentef, Michael/L-5717-2013},
	ORCID-Numbers = {Claassen, Martin/0000-0001-7580-0588 McIver, James/0000-0003-0182-5031 De la Torre, Alberto/0000-0002-6751-8205 Kennes, Dante/0000-0002-9838-6866 Gerber, Simon/0000-0002-5717-2626 Sentef, Michael/0000-0002-7946-0282},
	Unique-ID = {WOS:000707512400001},
}

@article{Sell2008OL,
	Author = {Sell, Alexander and Leitenstorfer, Alfred and Huber, Rupert},
	Title = "{Phase-locked generation and field-resolved detection of widely tunable terahertz pulses with amplitudes exceeding 100 MV/cm}",
	Journal = {Opt. Lett.},
	Year = {2008},
	Volume = {33},
	Number = {23},
	Pages = {2767-2769},
	Month = {DEC 1},
	url = {https://doi.org/10.1364/OL.33.002767},
	ISSN = {0146-9592},
	EISSN = {1539-4794},
	ResearcherID-Numbers = {Huber, Rupert/N-4126-2018 Leitenstorfer, Alfred/B-9561-2015},
	ORCID-Numbers = {Huber, Rupert/0000-0001-6617-9283 Leitenstorfer, Alfred/0000-0002-9847-257X},
	Unique-ID = {WOS:000262206500015},
}

@article{Liu2017OL,
	Author = {Liu, B. and Bromberger, H. and Cartella, A. and Gebert, T. and Foerst, M. and Cavalleri, A.},
	Title = "{Generation of narrowband, high-intensity, carrier-envelope phase-stable pulses tunable between 4 and 18 THz}",
	Journal = {Opt. Lett.},
	Year = {2017},
	Volume = {42},
	Number = {1},
	Pages = {129-131},
	Month = {JAN 1},
	url = {https://doi.org/10.1364/OL.42.000129},
	ISSN = {0146-9592},
	EISSN = {1539-4794},
	ResearcherID-Numbers = {Först, Michael/D-8924-2012 Cartella, Andrea/E-6099-2012},
	ORCID-Numbers = {Först, Michael/0000-0002-8057-4826 Liu, Biaolong/0000-0002-0505-605X},
	Unique-ID = {WOS:000391396800034},
}

@article{Lin2024ACSPho,
	Author = {Lin, Tong and Xu, Rui and Chen, Xiaotong and Guan, Yuxuan and Yao, Mingxing and Zhang, Junhao and Li, Xinwei and Zhu, Hanyu},
	Title = "{Subwavelength, Phase-Sensitive Microscopy of Third-Order Nonlinearity in Terahertz Frequencies}",
	Journal = {ACS Photonics},
	Year = {2023},
	Volume = {11},
	Number = {1},
	Pages = {33-41},
	Month = {NOV 15},
	url = {https://doi.org/10.1021/acsphotonics.3c00787},
	ISSN = {2330-4022},
	ResearcherID-Numbers = {Zhu, Hanyu/GZA-4788-2022 Li, Xinwei/KYR-0362-2024 xu, rui/GRX-5734-2022},
	ORCID-Numbers = {Lin, Tong/0009-0006-2147-4573 Zhu, Hanyu/0000-0003-3376-5352},
	Unique-ID = {WOS:001144581300001},
}

@article{Li2019Science,
	Author = {Li, Xian and Qiu, Tian and Zhang, Jiahao and Baldini, Edoardo and Lu, Jian and Rappe, Andrew M. and Nelson, Keith A.},
	Title = "{Terahertz field-induced ferroelectricity in quantum paraelectric SrTiO$_{3}$}",
	Journal = {Science},
	Year = {2019},
	Volume = {364},
	Number = {6445, SI},
	Pages = {1079},
	Month = {JUN 14},
	url = {https://doi.org/10.1126/science.aaw4913},
	ISSN = {0036-8075},
	EISSN = {1095-9203},
	ResearcherID-Numbers = {Baldini, Edoardo/HTM-2242-2023 Rappe, Andrew/HFZ-8855-2022 Lu, Jian/I-7738-2019},
	ORCID-Numbers = {Baldini, Edoardo/0000-0002-8131-9974 Qiu, Tian/0000-0001-8510-894X Zhang, Jiahao/0000-0002-8284-8122},
	Unique-ID = {WOS:000471306700043},
}

@article{Gao2024PtBi2,
	Author = {Gao, Y. and Zeng, X. Y. and Wang, X. B. and Shi, Y. G. and Cheng, L. and Qi, J.},
	Title = "{Terahertz manipulation of nonlinear optical response in topological material PtBi$_{2}$}",
	Journal = {Opt. Lett.},
	Year = {2024},
	Volume = {49},
	Number = {14},
	Pages = {3862-3865},
	Month = {JUL 15},
	url = {https://doi.org/10.1364/OL.525200},
	ISSN = {0146-9592},
	EISSN = {1539-4794},
	ResearcherID-Numbers = {Shi, Youguo/B-6316-2018},
	ORCID-Numbers = {Cheng, Liang/0000-0002-3514-2207 Wang, Xinbo/0009-0001-0565-3911},
	Unique-ID = {WOS:001290097900008},
}

@article{SrTiO3Arxiv2024,
	title="{A Hidden Quantum Paraelectric Phase in SrTiO$_{3}$ Induced by Terahertz Field}", 
	author = {Wei Li and Hanbyul Kim and Xinbo Wang and Jianlin Luo and Simone Latini and Dongbin Shin and Jun-Ming Liu and Jing-Feng Li and Angel Rubio and Ce-Wen Nan and Qian Li},
	year={2024},
	Journal = {arXiv:2412.20887},
	url={https://arxiv.org/abs/2412.20887}, 
}

@article{Yang2024PRB,
	title = "{Ultrafast carrier and phonon dynamics in ${\mathrm{Bi}}_{2}{\mathrm{Se}}_{3}$ under high pressure}",
	author = {Yang, Y. and Meng, Y. H. and Lu, B. R. and Jin, F. and Shi, Y. G. and Hong, F. and Zhang, S. S. and Yu, X. H. and Wang, X. B. and Luo, J. L.},
	journal = {Phys. Rev. B},
	volume = {109},
	issue = {6},
	pages = {064307},
	numpages = {8},
	year = {2024},
	month = {Feb},
	publisher = {American Physical Society},
	url = {https://doi.org/10.1103/PhysRevB.109.064307}
}

@article{Meng2024NatCommun,
	Author = {Meng, Yanghao and Yang, Yi and Sun, Hualei and Zhang, Sasa and Luo, Jianlin and Chen, Liucheng and Ma, Xiaoli and Wang, Meng and Hong, Fang and Wang, Xinbo and Yu, Xiaohui},
	Title = "{Density-wave-like gap evolution in La$_{3}$Ni$_{2}$O$_{7}$ under high pressure revealed by ultrafast optical spectroscopy}",
	Journal = {Nat. Commun.},
	Year = {2024},
	Volume = {15},
	Number = {1},
        Pages = {10408},
	Month = {NOV 29},
	url = {https://doi.org/10.1038/s41467-024-54518-1},
	Article-Number = {10408},
	EISSN = {2041-1723},
	ResearcherID-Numbers = {Ma, Xiaoli/HZI-7536-2023 Luo, Jianlin/CAA-7649-2022 Chen, Liu-Cheng/AAO-2409-2021 Hong, Fang/C-6070-2014 Wang, Meng/E-6595-2012},
	ORCID-Numbers = {Ma, Xiaoli/0000-0002-8835-4684 zhang, sasa/0000-0002-0568-3803 Yang, Yi/0009-0008-6147-947X Wang, Xinbo/0009-0001-0565-3911 Hong, Fang/0000-0003-0060-2063 Wang, Meng/0000-0002-8232-2331},
	Unique-ID = {WOS:001367893700039},
}

@article{Dong2018CPB,
	Author = {Dong, T. and Chen, Z. G. and Wang, N. L.},
	Title = "{Magneto optics and time resolved terahertz spectrocopy}",
	Journal = {Chin. Phys. B},
	Year = {2018},
	Volume = {27},
	Number = {7},
        Pages = {077501},
	Month = {JUL},
	url = {https://doi.org/10.1088/1674-1056/27/7/077501},
	Article-Number = {077501},
	ISSN = {1674-1056},
	EISSN = {1741-4199},
	ResearcherID-Numbers = {Wang, Nanlin/HPG-1600-2023 Chen, Zhiguo/B-9192-2015},
	ORCID-Numbers = {Chen, Zhiguo/0000-0002-8242-4784},
	Unique-ID = {WOS:000439128200001},
}

@article{Li2025Skyrmions,
	Author = {Li, Wei and Wang, Sixu and Peng, Pai and Han, Haojie and Wang, Xinbo and Ma, Jing and Luo, Jianlin and Liu, Jun-Ming and Li, Jing-Feng and Nan, Ce-Wen and Li, Qian},
	Title = "{Terahertz excitation of collective dynamics of polar skyrmions over a broad temperature range}",
	Journal = {Nat. Phys.},
	Year = {2025},
	Volume = {21},
	Number = {12},
        Pages = {1965-1972},
	Month = {DEC},
	url = {https://doi.org/10.1038/s41567-025-03056-8},
	EarlyAccessDate = {OCT 2025},
	ISSN = {1745-2473},
	EISSN = {1745-2481},
	ResearcherID-Numbers = {Li, Jing-Feng/D-2770-2014 Han, Haojie/OTK-0833-2025},
	ORCID-Numbers = {Liu, Jun-Ming/0000-0001-8988-8429 Li, Jing-Feng/0000-0002-0185-0512 Wang, Xinbo/0009-0001-0565-3911 A, AA/0009-0003-4574-2091},
	Unique-ID = {WOS:001586234900001},
}

@article{Huang2026NRP,
	Author = {Huang, Chuankun and Mootz, Martin and Luo, Liang and Perakis, Ilias E. and Wang, Jigang},
	Title = "{Terahertz 2D coherent spectroscopy for probing and controlling multicorrelations in quantum matter}",
	Journal = {Nat. Rev. Phys.},
	Year = {2026},
	Volume = {8},
	Number = {3},
	Pages = {171-185},
	Month = {MAR},
	url = {https://doi.org/10.1038/s42254-025-00917-2},
	EarlyAccessDate = {FEB 2026},
	EISSN = {2522-5820},
	ResearcherID-Numbers = {Perakis, Ilias/ABC-5659-2020 Luo, Liang/AAL-8437-2021},
	ORCID-Numbers = {Mootz, Martin/0000-0003-2040-3598 WANG, JIGANG/0000-0002-6159-4119},
	Unique-ID = {WOS:001681883100001},
}

@article{Fausti2011,
	Author = {Fausti, D. and Tobey, R. I. and Dean, N. and Kaiser, S. and Dienst, A. and Hoffmann, M. C. and Pyon, S. and Takayama, T. and Takagi, H. and Cavalleri, A.},
	Title = "{Light-Induced Superconductivity in a Stripe-Ordered Cuprate}",
	Journal = {Science},
	Year = {2011},
	Volume = {331},
	Number = {6014},
	Pages = {189-191},
	Month = {JAN 14},
	url = {https://doi.org/10.1126/science.1197294},
	ISSN = {0036-8075},
	EISSN = {1095-9203},
	ResearcherID-Numbers = {Hoffmann, Matthias/B-3893-2009 Hoffmann, Matthias/B-3893-2009 Pyon, Sunseng/B-2618-2011 TAKAGI, HIDENORI/Q-1041-2019 Kaiser, Stefan/B-7788-2008 Takayama, Tomohiro/LIC-1987-2024},
	ORCID-Numbers = {Dean, Nicky/0000-0002-4219-8807 Hoffmann, Matthias/0000-0002-3596-9853 Pyon, Sunseng/0000-0002-5716-1791 Kaiser, Stefan/0000-0001-9862-2788 Fausti, Daniele/0000-0002-2142-9741 Takayama, Tomohiro/0000-0002-3492-2025},
	Unique-ID = {WOS:000286433100035},
}

@article{Rowe2023,
	Author = {Rowe, E. and Yuan, B. and Buzzi, M. and Jotzu, G. and Zhu, Y. and Fechner, M. and F\"{o}rst, M. and Liu, B. and Pontiroli, D. and Ricc\`{o}, M. and Cavalleri, A.},
	Title = "{Resonant enhancement of photo-induced superconductivity in {K$_{3}$C$_{60}$}}",
	Journal = {Nat. Phys.},
	Year = {2023},
	Volume = {19},
	Number = {12},
	Pages = {1821},
	Month = {DEC},
	url = {https://doi.org/10.1038/s41567-023-02235-9},
	EarlyAccessDate = {OCT 2023},
	ISSN = {1745-2473},
	EISSN = {1745-2481},
	ResearcherID-Numbers = {Ricco, Mauro/D-9376-2017 Först, Michael/D-8924-2012 Pontiroli, Daniele/A-4543-2017 Fechner, Michael/J-7198-2018},
	ORCID-Numbers = {Ricco, Mauro/0000-0002-6879-2687 Först, Michael/0000-0002-8057-4826 Yuan, Bo/0000-0003-4721-0382 Pontiroli, Daniele/0000-0002-9990-539X Buzzi, Michele/0000-0001-7325-4632 Fechner, Michael/0000-0003-2774-7684 Jotzu, Gregor/0000-0003-4421-5874},
	Unique-ID = {WOS:001081652100003},
}

@article{Fava2024,
	Author = {Fava, S. and De Vecchi, G. and Jotzu, G. and Buzzi, M. and Gebert, T. and Liu, Y. and Keimer, B. and Cavalleri, A.},
	Title = "{Magnetic field expulsion in optically driven YBa$_{2}$Cu$_{3}$O$_{6.48}$}",
	Journal = {Nature},
	Year = {2024},
	Volume = {632},
	Number = {8023},
	Pages = {75},
	Month = {AUG 1},
	url = {https://doi.org/10.1038/s41586-024-07635-2},
	EarlyAccessDate = {JUL 2024},
	ISSN = {0028-0836},
	EISSN = {1476-4687},
	ORCID-Numbers = {Jotzu, Gregor/0000-0003-4421-5874 Keimer, Bernhard/0000-0001-5220-9023 Buzzi, Michele/0000-0001-7325-4632 Cavalleri, Anea/0000-0002-3143-0850 Liu, Yiran/0000-0002-7266-5437 Fava, Sebastian/0000-0001-8981-4120 Gebert, Thomas/0000-0002-6031-7689},
	Unique-ID = {WOS:001281636500001},
}

@article{Schlauderer2019,
	Author = {Schlauderer, S. and Lange, C. and Baierl, S. and Ebnet, T. and Schmid, C. P. and Valovcin, D. C. and Zvezdin, A. K. and Kimel, A. V. and Mikhaylovskiy, R. V. and Huber, R.},
	Title = "{Temporal and spectral fingerprints of ultrafast all-coherent spin switching}",
	Journal = {Nature},
	Year = {2019},
	Volume = {569},
	Number = {7756},
	Pages = {383},
	Month = {MAY 16},
	url = {https://doi.org/10.1038/s41586-019-1174-7},
	ISSN = {0028-0836},
	EISSN = {1476-4687},
	ResearcherID-Numbers = {Kimel, Alexey/D-5112-2012 Zvezdin, Anatoly/ABG-9179-2020 Lange, Christoph/KFR-2605-2024 Mikhaylovskiy, Rostislav/H-3053-2012 Huber, Rupert/N-4126-2018},
	ORCID-Numbers = {Zvezdin, Anatoly/0000-0002-6039-780X Lange, Christoph/0000-0002-2134-6612 Mikhaylovskiy, Rostislav/0000-0003-3780-0872 Huber, Rupert/0000-0001-6617-9283},
	Unique-ID = {WOS:000468123700034},
}

@article{Zhang2024,
	Author = {Zhang, Zhenya and Kanega, Minoru and Maruyama, Kei and Kurihara, Takayuki and Nakajima, Makoto and Tachizaki, Takehiro and Sato, Masahiro and Kanemitsu, Yoshihiko and Hirori, Hideki},
	Title = "{Spin switching in {Sm$_{0.7}$Er$_{0.3}$FeO$_{3}$} triggered by terahertz magnetic-field pulses}",
	Journal = {Nat. Mater.},
	Year = {2025},
	Volume = {24},
	Number = {2},
	pages={219},
	Month = {FEB},
	url = {https://doi.org/10.1038/s41563-024-02034-4},
	EarlyAccessDate = {OCT 2024},
	ISSN = {1476-1122},
	EISSN = {1476-4660},
	ResearcherID-Numbers = {Kanemitsu, Yoshihiko/D-2006-2014 Tachizaki, Takehiro/AGP-0381-2022 Hirori, Hideki/F-3027-2018 Sato, Masahiro/E-6806-2013},
	ORCID-Numbers = {Tachizaki, Takehiro/0000-0003-2708-2926 Hirori, Hideki/0000-0001-6056-8675 Kanega, Minoru/0009-0008-4623-8010 Kurihara, Takayuki/0000-0001-6903-002X},
	Unique-ID = {WOS:001341168600001},
}

@article{Tokura2017,
	Author = {Tokura, Yoshinori and Kawasaki, Masashi and Nagaosa, Naoto},
	Title = "{Emergent functions of quantum materials}",
	Journal = {Nat. Phys.},
	Year = {2017},
	Volume = {13},
	Number = {11},
	Pages = {1056-1068},
	Month = {NOV},
	url = {https://doi.org/10.1038/nphys4274},
	ISSN = {1745-2473},
	EISSN = {1745-2481},
	ResearcherID-Numbers = {Tokura, Yoshinori/C-7352-2009 Kawasaki, Masashi/B-5826-2008 Nagaosa, Naoto/G-7057-2012},
	ORCID-Numbers = {Tokura, Yoshinori/0000-0002-2732-4983 Kawasaki, Masashi/0000-0001-6397-4812},
	Unique-ID = {WOS:000414250500016},
}

@article{Mao2018,
	Author = {Mao, Ho-Kwang and Chen, Xiao-Jia and Ding, Yang and Li, Bing and Wang, Lin},
	Title = "{Solids, liquids, and gases under high pressure}",
	Journal = {Rev. Mod. Phys.},
	Year = {2018},
	Volume = {90},
	Number = {1},
        pages = {015007},
        numpages = {55},
	Month = {MAR 20},
	url = {https://doi.org/10.1103/RevModPhys.90.015007},
	Article-Number = {015007},
	ISSN = {0034-6861},
	EISSN = {1539-0756},
	ResearcherID-Numbers = {Chen, Xiaojia/JYV-2395-2024 Wang, Lin/N-4161-2017},
	Unique-ID = {WOS:000427820900001},
}

@article{Yamamoto2015,
	Author = {Yamamoto, Ayako and Takeshita, Nao and Terakura, Chieko and Tokura, Yoshinori},
	Title = "{High pressure effects revisited for the cuprate superconductor family with highest critical temperature}",
	Journal = {Nat. Commun.},
	Year = {2015},
	Volume = {6},
        Pages = {8990},
        Number = {1},
	Month = {DEC},
	url = {https://doi.org/10.1038/ncomms9990},
	Article-Number = {8990},
	ISSN = {2041-1723},
	ResearcherID-Numbers = {Terakura, Chieko/A-5959-2009 TAKESHITA, Nao/A-2948-2013 Tokura, Yoshinori/C-7352-2009},
	ORCID-Numbers = {Terakura, Chieko/0009-0002-8768-868X TAKESHITA, Nao/0000-0001-5081-2777 Tokura, Yoshinori/0000-0002-2732-4983},
	Unique-ID = {WOS:000367579000002},
}

@article{Dong2023,
	Author = {Dong, Tao and Zhang, Si-Jie and Wang, Nan-Lin},
	Title = "{Recent Development of Ultrafast Optical Characterizations for Quantum Materials}",
	Journal = {Adv. Mater.},
	Year = {2023},
	Volume = {35},
	Number = {27},
	Month = {JUL},
        Pages = {2110068},
	url = {https://doi.org/10.1002/adma.202110068},
	EarlyAccessDate = {NOV 2022},
	ISSN = {0935-9648},
	EISSN = {1521-4095},
	ResearcherID-Numbers = {Wang, Nan-Lin/HPG-1600-2023 Zhang, Sijie/K-8204-2018},
	ORCID-Numbers = {Wang, Nan-Lin/0000-0003-2584-9265},
	Unique-ID = {WOS:000890449000001},
}

@article{Chen2023ZrN,
	Author = {Chen, Fucong and Bai, Xinbo and Wang, Yuxin and Dong, Tao and Shi, Jinan and Zhang, Yanmin and Sun, Xiaomin and Wei, Zhongxu and Qin, Mingyang and Yuan, Jie and Chen, Qihong and Wang, Xinbo and Wang, Xu and Zhu, Beiyi and Huang, Rongjin and Jiang, Kun and Zhou, Wu and Wang, Nanlin and Hu, Jiangping and Li, Yangmu and Jin, Kui and Zhao, Zhongxian},
	Title = "{Emergence of superconducting dome in ZrN$_{x}$ films via variation of nitrogen concentration}",
	Journal = {Sci. Bull.},
	Year = {2023},
	Volume = {68},
	Number = {7},
	Pages = {674-678},
	Month = {APR 15},
	url = {https://doi.org/10.1016/j.scib.2023.03.018},
	EarlyAccessDate = {APR 2023},
	ISSN = {2095-9273},
	EISSN = {2095-9281},
	ResearcherID-Numbers = {Qin, Mingyang/LRT-2659-2024 Jin, Kui/AAT-4728-2021 Bai, Xinbo/OFO-4317-2025 Wang, Nanlin/HPG-1600-2023 Zhou, Wu/D-8526-2011 hu, jiangping/C-3320-2014 Yuan, Jie/D-4110-2014 Huang, Rong/OZE-0833-2025 Zhu, Bei/C-1506-2011 Li, Yangmu/AAW-6716-2020 Wang, Yuxin/J-2050-2015},
	ORCID-Numbers = {Qin, Mingyang/0000-0001-5341-7465 Wang, Xinbo/0009-0001-0565-3911 Zhou, Wu/0000-0002-6803-1095},
	Unique-ID = {WOS:000986496100001},
}

@article{Duvillaret1996JSTQE,
	Author = {Duvillaret, L and Garet, F and Coutaz, JL},
	Title = "{A reliable method for extraction of material parameters in terahertz time-domain spectroscopy}",
	Journal = {IEEE J. Sel. Top. Quantum Electron.},
	Year = {1996},
	Volume = {2},
	Number = {3},
	Pages = {739-746},
	Month = {SEP},
	url = {https://doi.org/10.1109/2944.571775},
	ISSN = {1077-260X},
	Unique-ID = {WOS:A1996WX09700034},
}

@article{Nicoletti2016AOP,
	Author = {Nicoletti, Daniele and Cavalleri, Andrea},
	Title = "{Nonlinear light-matter interaction at terahertz frequencies}",
	Journal = {Adv. Opt. Photonics},
	Year = {2016},
	Volume = {8},
	Number = {3},
	Pages = {401-464},
	Month = {SEP 30},
	url = {https://doi.org/10.1364/AOP.8.000401},
	ISSN = {1943-8206},
	Unique-ID = {WOS:000384831500002},
}

@article{Hebling2008JOSAB,
	Author = {Hebling, Janos and Yeh, Ka-Lo and Hoffmann, Matthias C. and Bartal, Balazs and Nelson, Keith A.},
	Title = "{Generation of high-power terahertz pulses by tilted-pulse-front excitation and their application possibilities}",
	Journal = {J. Opt. Soc. Am. B},
	Year = {2008},
	Volume = {25},
	Number = {7},
	Pages = {B6-B19},
	Month = {JUL},
	url = {https://doi.org/10.1364/JOSAB.25.0000B6},
	ISSN = {0740-3224},
	EISSN = {1520-8540},
	ResearcherID-Numbers = {Hoffmann, Matthias/B-3893-2009 Hebling, János/AAH-1737-2019 Hoffmann, Matthias/B-3893-2009},
	ORCID-Numbers = {Hebling, János/0000-0001-9669-2977 Hoffmann, Matthias/0000-0002-3596-9853},
	Unique-ID = {WOS:000258014000003},
}

@article{Fu2024DyNdFeO3,
	Author = {Fu, Zhichao and Chen, Junyu and Shang, Jiamin and Lin, Xian and Suo, Peng and Sun, Kaiwen and Wang, Chen and Li, Qixin and Luo, Jianlin and Wang, Xinbo and Wu, Anhua and Ma, Guohong},
	Title = "{Magnon and electromagnon excitations in Dy$_{0.9}$Nd$_{0.1}$FeO$_{3}$ single crystals tuned with temperature and magnetic field}",
	Journal = {Appl. Phys. Lett.},
	Year = {2024},
	Volume = {125},
	Number = {24},
       Pages = {241102},
	Month = {12},
	url = {https://doi.org/10.1063/5.0243672},
	Article-Number = {241102},
	ISSN = {0003-6951},
	EISSN = {1077-3118},
	ResearcherID-Numbers = {Wu, Anhua/AEM-1888-2022 Luo, Jianlin/CAA-7649-2022},
	ORCID-Numbers = {Wang, Xinbo/0009-0001-0565-3911 , Zhichao Fu/0009-0009-5266-2990 Lin, Xian/0000-0003-1114-4598 Sun, Kaiwen/0009-0001-8781-1177},
	Unique-ID = {WOS:001378199200005},
}

@article{Shi2025Na2Co2TeO6,	
	title = "{Field and temperature evolution of the magnetic excitations in the field-induced state of ${\mathrm{Na}}_{2}{\mathrm{Co}}_{2}{\mathrm{TeO}}_{6}$}",
	author = {Shi, Liyu and Li, Xintong and Li, Rongsheng and Li, Yuan and Dong, Tao and Luo, Jianlin and Wang, Xinbo and Wang, Nanlin},
	journal = {Phys. Rev. B},
	volume = {112},
	issue = {18},
	pages = {184406},
	numpages = {8},
	year = {2025},
	month = {Nov},
	publisher = {American Physical Society},
	url = {https://doi.org/10.1103/6s3y-47gv}
}

@article{Chen2025GdHoFeO3,
	Author = {Chen, Junyu and Li, Qixin and Fu, Zhichao and Shang, Jiamin and Suo, Peng and Lin, Xian and Luo, Jianlin and Wang, Xinbo and Wu, Anhua and Ma, Guohong},
	Title = "{Terahertz cavity magnon-polaritons in Gd$_{0.5}$Ho$_{0.5}$FeO$_{3}$ single crystals tuned with temperature and magnetic field}",
	Journal = {Appl. Phys. Lett.},
	Year = {2025},
	Volume = {127},
	Number = {6},
        Pages = {062402},
	Month = {AUG 11},
	url = {https://doi.org/10.1063/5.0277632},
	Article-Number = {062402},
	ISSN = {0003-6951},
	EISSN = {1077-3118},
	ResearcherID-Numbers = {Wu, Anhua/AEM-1888-2022},
	ORCID-Numbers = {Ma, Guo-Hong/0000-0002-3972-5012 , JunYu Chen/0009-0009-9732-2076 Wu, Anhua/0000-0002-0979-8085 Lin, Xian/0000-0003-1114-4598 Wang, Xinbo/0009-0001-0565-3911 , Zhichao Fu/0009-0009-5266-2990},
	Unique-ID = {WOS:001548100700001},
}

@article{Li2026GdHoFeO3,
	Author = {Li, Qixin and Shen, Hui and Luo, Jianlin and Wu, Anhua and Wang, Xinbo and Kalashnikova, Alexandra M. and Xu, Jiayue and Su, Liangbi},
	Title = "{Anomalous magnetic field-dependent spin dynamics in Gd$_{0.3}$Ho$_{0.7}$FeO$_{3}$ single crystal near the ordering temperature of rare-earth ions}",
	Journal = {Appl. Phys. Lett.},
	Year = {2026},
	Volume = {128},
	Number = {8},
        Pages = {082408},
	Month = {FEB 23},
	url = {https://doi.org/10.1063/5.0296964},
	Article-Number = {082408},
	ISSN = {0003-6951},
	EISSN = {1077-3118},
	ResearcherID-Numbers = {Kalashnikova, Alexandra/C-7821-2014},
	Unique-ID = {WOS:001702945400001},
}
	

\end{document}